\def\cm2{cm$^{-2}$}

\def\c2{C~{\sc ii}}
\def\c4{C~{\sc iv}}
\def\fe2{Fe~{\sc ii}}
\def\fe3{Fe~{\sc iii}}
\def\mg1{Mg~{\sc i}}
\def\mg2{Mg~{\sc ii}}
\def\si2{Si~{\sc ii}}
\def\si4{Si~{\sc iv}}
\def\al2{Al~{\sc ii}}
\def\al3{Al~{\sc iii}}
\def\o1{O~{\sc i}}
\def\n1{N~{\sc i}}
\def\h1{H~{\sc i}}

\def\approxlt{\mathrel{\spose{\lower 3pt\hbox{$\sim$}}
        \raise 2.0pt\hbox{$<$}}}
\def\approxgt{\mathrel{\spose{\lower 3pt\hbox{$\sim$}}
        \raise 2.0pt\hbox{$>$}}}

\documentclass[article]{has40}
\usepackage{graphicx}
\usepackage{amssymb}
\tabletypesize{\normalsize}  

\def\plotone#1{\centering \leavevmode
\includegraphics[width=.95\columnwidth]{#1}}

\def\plotone#1{\centering \leavevmode
\includegraphics[width=.95\columnwidth]{#1}}

\shortauthors{Van Noord et al.}
\shorttitle{Automated searches for variable stars}

\begin{document}
\large    
\pagenumbering{arabic}
\setcounter{page}{38}

\title{Automated searches for variable stars}

%
%
\author{{\noindent  Daniel M. Van Noord{$^{\rm 1}$}, Lawrence A. Molnar{$^{\rm 1}$}, and Steven D. Steenwyk{$^{\rm 1}$}\\
\\
{\it (1) Department of Physics and Astronomy, Calvin College, Grand Rapids, MI, USA} 
}
}

%
%
\email{(1) dmv24@students.calvin.edu}


\begin{abstract}
With recent developments in imaging and computer technology the amount of available astronomical data has increased dramatically. Although most of these data sets are not dedicated to the study of variable stars much of it can, with the application of proper software tools, be recycled for the discovery of new variable stars. Fits Viewer and Data Retrieval System is a new software package that takes advantage of modern computer advances to search astronomical data for new variable stars. More than 200 new variable stars have been found in a data set taken with the Calvin College Rehoboth Robotic telescope using FVDRS. One particularly interesting example is a very fast subdwarf B with a 95 minute orbital period, the fastest currently known of the HW Vir type.
\end{abstract}

\section{Development history}
Development of Fits Viewer and Data Retrieval System (FVDRS) began around three years ago. The program is written entirely in Java. This design decision was made to take advantage of the multiplatform nature of the Java language. Although the discovery of new variable stars is the primary purpose of the program several different alternate versions have been created to solve various astronomical problems. These include programs to calculate sky transparency, colors of stars and asteroids, as well as several other less used programs. The most used version, Variable Star Search, uses numerous new automated finding algorithms to assist in the discovery of new variable stars.
\section{Capabilities}
Efficiency was a major development goal of the Variable Star Search program. The current version of the program is capable of processing up to 20 MB/s of images using 50\% of computer resources on an Intel i5 processor with 4GB of ram. The program is designed to scale up based on the computer that the code is on.  This means that the program is highly versatile, capable of analyzing various data sets of almost any scope. In addition, automatic multi-core setup results in increased ease of use.
\section{Applications}
So far FVDRS Variable Star Search has been tested on a data set taken at Calvin College using our 0.4 m telescope in Rehoboth, New Mexico.
The data set was obtained by Melissa Dykhuis as part of a study of the spin orientations of the asteroids in the Flora collisional family.
This makes it ideal for the discovery of short period variables for several reasons. The first reason is the area and time range covered by the survey. In total the asteroid data generated around 400 good fields for searching, covering 36 square degrees. A good field contains approximately 60 images spanning 4 hours. Each image covers an area of ~14 by 23 arc minutes and contains between 50 and 1000 detectable stars. The second reason was the depth of coverage, the limiting magnitudes tended to be between 17th and 19th magnitude. Finally the survey has little bias in position. Because the data followed the ecliptic the survey's galactic latitude and longitude was not chosen and as a result, variables of nearly all types have been discovered. The table shows variables found during the initial testing of four weeks. Several hundred additional variables have been found with the software after this initial test period.
For about half of these systems, we have obtained followup data (either from the Catalina Surveys database or taking new data with our telescope) sufficient to characterize the variable classification and period.
These have been published in the Variable Star Index.

In addition to the scientific version, a laboratory version has been created. This version has been successfully used from the middle school level through the introductory college level, allowing students to experience the excitement of their own new variable star discovery and engaging them in doing real science, hands on.
\begin{center}
\begin{deluxetable*}{rcccccc}
\tabletypesize{\normalsize}
\tablecaption{Variable Discoveries as of summer 2011}
\tablewidth{0pt}
\tablehead{ \\ \colhead{Variable Type}   & \colhead{Found in past at Calvin} & \colhead{Found during initial testing} } \\

\startdata
 RR Lyrea &  4  & 17 \\
 Delta Scuti &  0  & 3 \\
 Mira &  1  & 2 \\
 Semiregular &  0  & 16 \\
 Eclipsing Algol &  3  & 3 \\
 Eclipsing Binary &  4  & 3 \\
 W Ursa Majoris &  22  & 21 \\
 RS CVn &  0  & 2 \\
 Elliptical &  0  & 3 \\
 Unknown or other type &  2  & 42 \\
\enddata
\end{deluxetable*}
\end{center}
 
\section{Future work}
The next step in the development of the software is to apply it to a larger dataset. A larger survey will allow for continuing development of the automated finding algorithms. More data will allow for better weights to be calculated for the different finding algorithms. In addition to the limited single computer version, a new version is planned that will run online, allowing interested parties to upload and search data. Using archival data, institutions without their own telescope could run a variable star discovery lab with this online tool.
\section{V2008$-$1753}
V2008$-$1753 (see Figure 1) was one of the most interesting discoveries made using FVDRS. This binary system was discovered within the Calvin data archive and is a subdwarf B star and a main sequence M dwarf, a combination known as an HW Vir system. This system has the shortest period of any known HW Vir system, with a period that is less than 95 minutes.  The extremely short period may put constraints on the formation models for these systems. Further work is ongoing, with a paper in preparation giving more details of this fascinating system. The discovery of this system and several other interesting variable systems demonstrates the value of efficient searching of archival data.

\begin{figure*}
\centering
\plotone{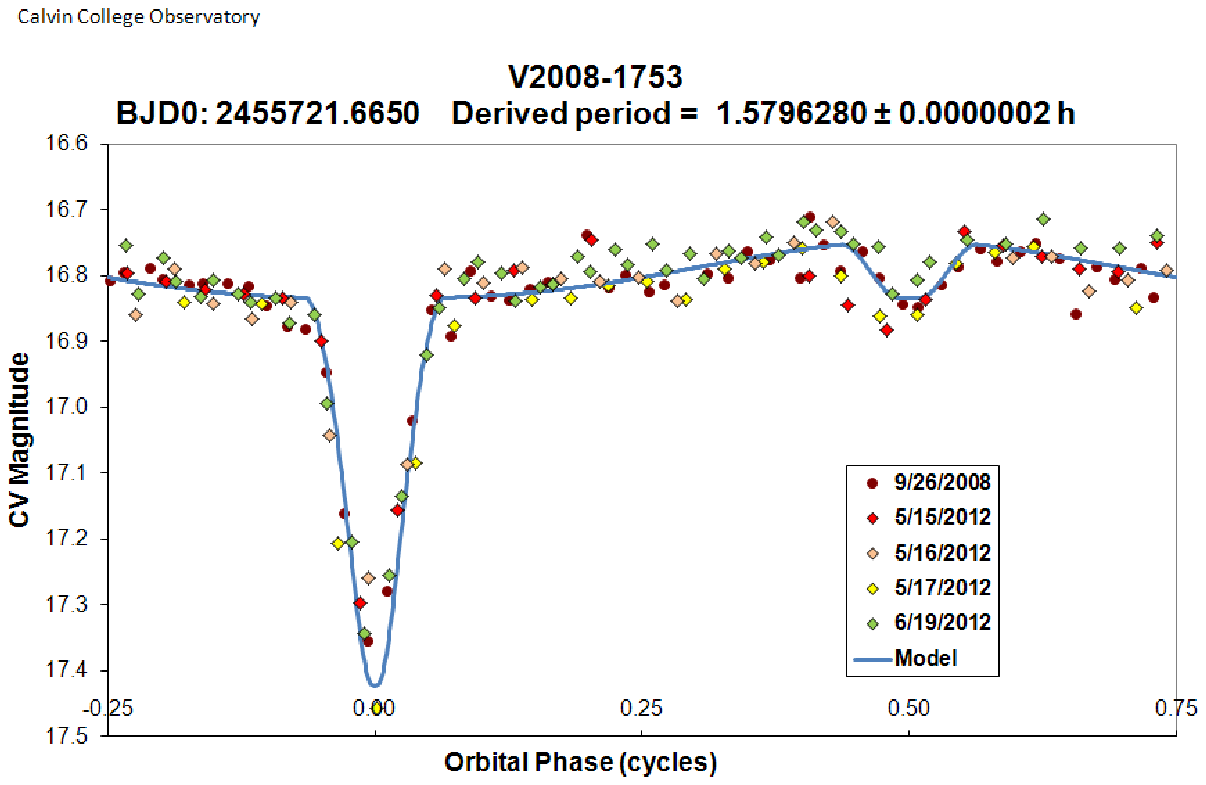}
\vskip0pt
\caption{This is the light curve of V2008$-$1753. The eclipse depth indicates that a very compact hot object is being eclipsed by a small cool object, one side of which is strongly heated by the other object. The increase in brightness leading up to the secondary eclipse is the reflection effect. The amplitude of the reflection and the color of the system indicate a subdwarf B star as the primary.}
\label{o1039}
\end{figure*}

\section{Conclusion}
Modern advances in computer technology have made it possible for high end data searching on simple home computers. FVDRS takes advantage of these new technologies to allow for simple data processing. Future versions will improve both the accuracy and the scope of this software. Through the mining of old data sets interesting new objects, such as V2008$-$1753, can be discovered, helping to fill in the current understanding of the universe without the need for more observations devoted to discovery.

\section{Acknowledgments}
DMV and SDS acknowledge support from the Michigan Space Grant Consortium.  DMV has also received support from the Calvin College Integrated Science Research Institute.

\end{document}